# Dimensionality reduction, and function approximation of poly(lactic-co-glycolic acid) micro- and nanoparticle dissolution rate

Varun Kumar Ojha[1,2]
Konrad Jackowski[3]
Ajith Abraham[1,4]
Václav Snášel[1,2]

[1]IT4Innovations, VŠB – Technical University of Ostrava, Ostrava, Czech Republic; [2]Department of Computer Science, VŠB – Technical University of Ostrava, Ostrava, Czech Republic; [3]Department of Systems and Computer Networks, Wrocław University of Technology, Wrocław, Poland; [4]Machine Intelligence Research Labs, Auburn, WA, USA

Correspondence: Varun Kumar Ojha
IT4Innovations, VŠB – Technical University of Ostrava, 17 listopadu 15/2172, 708 33 Ostrava – Poruba, Ostrava, Czech Republic
Tel +420 777 880431
Email varun.kumar.ojha@vsb.cz

**Abstract:** Prediction of poly(lactic-co-glycolic acid) (PLGA) micro- and nanoparticles' dissolution rates plays a significant role in pharmaceutical and medical industries. The prediction of PLGA dissolution rate is crucial for drug manufacturing. Therefore, a model that predicts the PLGA dissolution rate could be beneficial. PLGA dissolution is influenced by numerous factors (features), and counting the known features leads to a dataset with 300 features. This large number of features and high redundancy within the dataset makes the prediction task very difficult and inaccurate. In this study, dimensionality reduction techniques were applied in order to simplify the task and eliminate irrelevant and redundant features. A heterogeneous pool of several regression algorithms were independently tested and evaluated. In addition, several ensemble methods were tested in order to improve the accuracy of prediction. The empirical results revealed that the proposed evolutionary weighted ensemble method offered the lowest margin of error and significantly outperformed the individual algorithms and the other ensemble techniques.

## Introduction

Predicting the poly(lactic-co-glycolic acid) (PLGA) micro- and nanoparticle dissolution profiles presents a complex and vital problem. The complexity of the problem can be understood from the fact that academic literature[1–18] provides 300 potential factors that may influence the dissolution of the PLGA protein particles.[19] After analyzing the collected dataset, the primary approach adopted in most research has been to reduce the dimensionality of the dataset. Dimensionality reduction techniques transform high-dimensional datasets into low-dimensional datasets, thereby improving the model's computational speed, predictability, and generalization ability. Dimensionality reduction techniques are classified into two categories: feature selection, and feature extraction. The feature selection technique is useful when the available dataset has a large dimension and relatively few cases (samples), whereas the feature extraction technique is useful when the dataset has a large dimension and high redundancy.[20]

The dataset in the present research had a large dimension, and the features appeared to have high redundancy. Therefore, it was not immediately clear to us whether we should use feature selection or feature extraction. Hence, we explored both feature selection and feature extraction techniques in order to find the best possible solution. Several regression models were employed to evaluate the relationship between the obtained input variables (features) and output variable.

In the scope of the present study, our focus was on PLGA nano- and microsphere dissolution properties and drug release rate. Szlęk et al[21] and Fredenberg et al[22] described that drug release from the PLGA matrix is mainly governed by two mechanisms: diffusion, and degradation/erosion. Several factors influencing the diffusion and degradation rates of PLGA as described by Kang et al, [23,24] Blanco and Alonso,[25] and Mainardes and Evangelista[26] include pore diameters, matrix active pharmaceutical ingredient (API) interactions, API–API interactions, and the composition of the formulation. Szlęk et al[21] developed a predictive model to describe the underlying relationship between those influencing factors on the drug's release profile, and they focused on feature selection, artificial neural network, and genetic programming approaches to come up with a suitable prediction model. In the past, several mathematical models, including the Monte Carlo and cellular automata microscopic models, were proposed by Zygourakis and Markenscoff,[27] and Gopferich.[28] A partial differential equations model was proposed by Siepmann et al[29] to address the influence of underlying PLGA properties on the drug's release rate/protein dissolution.

The highlights of the present article are as follows:
- a comprehensive discussion on the drug release problem and dataset collection mechanisms;
- a comprehensive discussion on various computational tools used to reduce dimensionality of dataset;
- a concise discussion on the elementary regression models available in the literature;
- a concise discussion on the ensemble methods used for making ensembles of the elementary regression models;
- a comprehensive discussion and conclusion on the experimental results mentioned in the present article.

## Methodology
### A description of the problem
PLGA micro- and nanoparticles could play a significant role in the medical application and toxicity evaluation of PLGA-based multi-particulate dosages.[30] PLGA micro-particles are important diluents used to produce drugs in their correct dosage form. Apart from playing the role as a filler, PLGA as an excipient, and alongside pharmaceutical APIs, plays other crucial roles in various ways. It helps in the dissolution of drugs, thus increasing the absorbability and solubility of drugs.[31,32] It helps in pharmaceutical manufacturing processes by improving API powders' flow and non-stickiness.

The dataset collected from various academic literature[1–18] contains 300 input features categorized into four groups, including protein descriptor, plasticizer, formulation characteristics, and emulsifier. A detailed description of the dataset is given in Table 1. For example, the formulation characteristics group contains features such as PLGA-inherent viscosity, PLGA molecular weight, lactide-to-glycolide ratio, inner and outer phase polyvinyl alcohol (PVA) concentration, PVA molecular weight, inner phase volume, encapsulation rate, mean particle size, PLGA concentration, and experimental conditions (dissolution pH, the number of dissolution additives, dissolution additive concentration, and production method and dissolution time). The protein descriptor, plasticizer, and emulsifier feature groups contain 85, 98, and 101 features, respectively. The regression model sought to predict the dissolution percentage or solubility of PLGA, which is dependent on the features mentioned above. In order to avoid over-fitting, collected data were preprocessed by adding noise to them. The dataset was then normalized, in other words, scaled within the range –1.0 and 1.0.

### Dimensionality reduction
#### Feature selection tools
Feature selection techniques enable us to identify the most relevant input feature from the available set of input features and allows us to avoid expensive (both in time and cost) experimental examination while developing a prediction model.[33]

#### Backward feature elimination
Backward feature elimination filtering starts with the maximum number of features (in this case, it starts with 300 features) and eliminates them one-by-one in an iterative manner. At each iteration, the resulting accuracy of

**Table 1** The PLGA dataset description

| Sl No | Group name | No of features | Importance |
|---|---|---|---|
| 1 | Protein descriptors | 85 | Describes the type of molecules and proteins used |
| 2 | Formulation characteristics | 17 | Describe the molecular properties such as molecular weight, particle size, etc |
| 3 | Plasticizer | 98 | Describe the properties such as fluidity of the material used |
| 4 | Emulsifier | 99 | Describe the properties of stabilizing/increase the pharmaceutical product life |
| 5 | Time in days | 1 | Time taken to dissolve |
| 6 | % of molecules dissolved | 1 | Output |

**Abbreviations:** PLGA, poly(lactic-co-glycolic acid); Sl, serial; No, number.

prediction is evaluated for all combinations of the remaining attributes. The subsets of attributes with the high accuracies are propagated to the next iteration. Finally, the subset with the highest degree of accuracy (the lowest root mean square error [RMSE]) is selected as the best subset.

### Correlation-based feature selection
Correlation-based feature selection assesses the value of a group of attributes that concern the individual predictive ability of each feature, as well with the possibility of repetition among the features.[34]

### Classifier-based feature selection
Classifier-based feature selection evaluates attribute subsets on training data and uses a classifier to estimate the merits of a set of attributes. A search algorithm is then applied to search for a suitable feature from among all the available feature sets.

### Wrapper feature selection
Wrapper-based feature selection evaluates attribute sets by using a learning scheme, and then uses cross-validation (CV) to estimate the accuracy of the learning scheme for a particular set of attributes.[35] A search algorithm is then applied to search for a suitable feature set from among all the available feature sets.

### Feature extraction
When it is affordable to easily generate test features, feature extraction techniques may be useful for dimensionality reduction. A regression model with a reduced input dimension may perform as well as it can if it has a complete set of features.[20] Therefore, feature extraction helps in reducing the computational overhead that may be incurred when using a complete input dimension.

### Principle component analysis
Principle component analysis (PCA) is a linear dimensionality reduction technique that transforms correlated data into uncorrelated data in a reduced dimension by finding a linear basis of reduced dimensionality for data with maximal variance. More specifically, it transfers correlated variables into a set of linearly uncorrelated variables called principle components.[36,37]

### Factor analysis
Factor analysis (FA), as opposed to PCA, determines whether a number of features of interest are linearly related to a smaller/reduced number of newly-defined features called factors. In other words, it discovers a reduced number of relatively independent features by mapping correlated features to a small set of features known as factors.[38]

### Independent component analysis
Independent component analysis (ICA), proposed by Hyvärinen and Oja[39] and Hyvärinen,[40] is a linear dimension reduction technique that transforms multidimensional feature vectors into components that are statistically as independent as possible. More specifically, ICA maps the observed variables (features) to a small number of latent variables (features) that are non-Gaussian and mutually independent, and they are called the independent components of the observed data.[41]

### Kernel PCA
Kernel PCA (kPCA) is an extension of PCA that uses kernel methods. kPCA computes the principal eigenvectors of the kernel matrix, rather than those of the covariance matrix.[42] Reformulating PCA in the kernel space is straightforward, since a kernel matrix is similar to the inner product of the data points in the high-dimensional space that is constructed using the kernel function. Typically, Gaussian, tangent hyperbolic, polynomial, and other functions are used for the kernel.

### Multidimensional scaling
Multidimensional scaling (MDS) is a non-linear dimension reduction technique that maps high-dimensional data representation into a low-dimensional representation while retaining the pairwise distances between the data points as much as possible. More specifically, MDS is used to analyze similarities or proximities between pairs of data points.[43]

## Function approximation algorithms
A regression/prediction model tries to build the relationship between independent variables $X$ (input) and dependent variables $y$ (output).[44] Moreover, it tries to find unknown parameters $\beta$ such that the error (2) is minimized, given $t$ predicted output $\hat{y}$ as:

$$\hat{y} = f(X, \beta). \quad (1)$$

Let $e_i = (\hat{y}_i - y_i)$ be the difference between the dependent variable $y$ and the predicted value $\hat{y}$. Therefore, the RMSE $\xi$ over data samples of size $n$ may be given as:

$$\xi = \sqrt{\frac{1}{n} \sum_{i=1}^{n} e_i^2}. \quad (2)$$

Regression models such as linear regression (LReg), Gaussian process regression (GPReg), multilayer perceptron (MLP), and sequential minimal optimization regression (SMOReg) are as follows.

### LReg
LReg is the simplest predictive model where independent variables ($|X|=n\times p$), dependent variable ($|y|=p$), with noise/error ($|\varepsilon|=p$), may be written as:

$$y_i = \beta_1 x_i 1 + \beta_2 x_i 2 + \cdots + \beta_p x_i p = x_i^T \beta \varepsilon_i. \quad (3)$$

### GPReg
The GPReg described by Rasmussen and Williams[45] and Rasmussen and Nickisch[46] is easily identified by its mean function $m(x)$ and covariance function $k(x, x')$. This is a natural generalization of the Gaussian distribution, whose mean $m$ and covariance $k$ are a vector and a matrix, respectively. Gaussian distribution is defined over vectors, whereas the Gaussian process is defined over functions $f$. Therefore, we may write:

$$f \sim GP(m, k). \quad (4)$$

Considering a zero mean, linear and non-linear covariance functions may be given as:

$$k(x, x') = \alpha x^T x' + \gamma, \quad (5)$$

$$k(x, x') = \alpha \exp\left(-\frac{\gamma}{2}(x-x')^T(x-x')\right), \quad (6)$$

where $\alpha$ and $\gamma$ are the parameters of the basis function.

### MLP
MLP is a feed-forward neural network having one or more hidden layers in between the input and output layers.[47,48] A neuron in an MLP first computes a linear-weighted combination of real-valued inputs, and then limits its amplitude using a non-linear activation function. In the presented research, MLP was trained using the backpropagation algorithm[49] and the resilient propagator.[50]

### Reduced error pruning tree
Reduced error pruning (REP) tree is a fast decision tree learner. It builds a decision tree based on information gain or reduction of the variance and prunes it using reduced-error pruning with over-fitting.[51,52]

### SMOReg
Sequential minimal optimization (SMO), an algorithm for the training of support vector regression proposed by Smola and Schölkopf,[53,54] and Schölkopf and Burges,[55] is an extension of the SMO algorithm proposed by Platt[56] for the support vector machine classifier. The idea of support vector regression is based on the computation of a linear regression function in a high-dimensional feature space where the input data are mapped using a non-linear function; support vector regression tries to minimize the generalization error in order to achieve generalized performance.

## The ensemble of function approximators
Getting the best regression algorithm is not a trivial task. Apart from having a plethora of options as listed in the present section, one has to decide what the optimal sets of parameters for each algorithm are. There is generally very little guidance available to address the question of how to select an algorithm and adjust its parameters for a specific problem. In such cases, experimental tests can help the user to make decisions. Still, in many cases the obtained results are not satisfactory or even not acceptable. In such situations, the ensemble approach can be used. Basically, it relies on the assumption that the properly-modeled fusion of responses of several elementary predictors will produce more accurate results and reduce the regression error.[57] Formally, let $\prod$ be a set of $k$ predictors given as:

$$\Pi = \{f_1, f_2, \cdots, f_k\}, \quad (7)$$

where $f_k$ indicates the state of the $k$th predictor. Each of the predictors is trained independently. The ensemble system fuses the outputs produced by the predictors in set $\prod$. In the simplest form, the ensemble can take the form of a simple average called the mean output regression, given as:

$$F'(x) = \frac{1}{k} \sum_{i=1}^{k} f_i(x), \quad (8)$$

where $F'$ is an ensemble system. The natural advantage of this model is its simplicity, since the output of the ensemble can easily be obtained by simple mathematical transformation without the necessity of setting any additional parameters. On the other hand, the main drawback of this model is that it treats all the elementary predictors as equally important, regardless of their quality. Weak predictors affect the final output to the same degree as strong ones. As a result, the quality of the ensemble is close to the average of all its constituents. Better results can be obtained when the contribution of a particular predictor depends on its quality. The greater the accuracy of the predictor, the greater its weight in the ensemble. The ensemble method is therefore called the quality weighted output regression, given as:

$$F'(x) = \frac{1}{k}\sum_{i=1}^{k} w_i f_i(x), \qquad (9)$$

where $\sum_{i=1}^{k} w_i = 1$. In its simplest form, the weights should be counter-proportional to the RMSE of the given predictors. However, in more advanced algorithms, the weights can be set over the course of time, eg, be an application of evolutionary algorithms.

## Diversity of the ensemble

There are several issues that have to be dealt with in order to make the application of the ensemble approach effective. One of the most essential issues is maintaining diversity among the predictors in the ensemble. Collecting a set of similar regression algorithms does not allow users to take any advantage from their fusion. Diversity can be ensured by applying one of the following procedures:
1. collecting predictors based on different models;
2. differentiating elementary predictor inputs.

In the first approach, it is assumed that different regression algorithms naturally make errors that are uncorrelated, even when they are trained on the same data. The second group consists of algorithms that create an ensemble based on the same regression model, but diversity is caused by training each of them on data partitions (as diversity occurred due to data partition in the Bagging algorithm) or using heterogeneous feature sets (the techniques used in random subspace [RS] algorithms).

## RS algorithms

RS is a method of constructing an ensemble of predictors where a pseudorandom procedure is used to select components of a feature vector separately for each ensemble constituent. The output of the ensemble is then obtained by averaging the outputs.[58]

## Bagging algorithms

Breiman[59] introduced the bagging method, which is basically a combination of multiple predictors. At first, subsets are prepared by cutting the original dataset using bootstrapping. A sequence of predictors is then allowed to run over the subsets of the dataset. Finally, the results from each of the predictors are aggregated using voting in order to get the final results. This method is supposed to enhance the performance of ensemble systems and reduce variances in order to improve predictability.[60,61]

## The evolutionary weighted ensemble

The evolutionary weighted ensemble (EWE) is used to make decisions, based on Equation 8. The learning process searches for a set of weight that minimizes the RMSE of the ensemble, and for that purpose, the learning set is used. Therefore, the objective function for the learning procedure or the ensemble system can be written as:

$$RMSE^{F'}(w_1, w_2, \cdots, w_k) = \sqrt{\frac{1}{N}\sum_{i}^{N}\sum_{j}^{k} w_j f_j(x_i - y_i)^2}, \quad (10)$$

where $x_i$ and $y_i$ denote the $i$th input–target pair in the learning set that consists of a total of $N$ samples.

We used the evolutionary algorithm,[62] which processes a population of possible solutions encoded as chromosomes. An overview of the EWE training procedure is presented in Figure 1. The components of the EWE algorithms are defined as follows:

### Initial population

The first step in the learning algorithm is generating an initial population. This consists of an arbitrarily chosen number of individuals with randomly selected weights that are scaled in order to ensure that their sum is 1.

### Evaluation of the population

Each individual is evaluated using an objective function. Obtained values determine the further behavior of the algorithm, especially selection procedures.

### Selection of the elite

The stability of the learning procedure is maintained by selecting two individuals with the smallest RMSE values. Those

---

Input: learning set
S – Population size
G – The number of generations
Π – A set of individual predictors
*repeat*
  Initialize population
  *for* t = 1 to G *do*
    Evaluate population over learning set
    Select elite
    Select parents
    Mutation
    Crossover
    Create offspring of the population
  *end* for
*until* the stopping criterion satisfied.

**Figure 1** Evolutionary weighted ensemble algorithm.

individuals, called the elite, are not affected by mutation or crossover operators and join the offspring population.

### Selection of the parents

Only selected individuals participate in generating offspring for the new generation population. The selection is based on their fitness and is done in a probabilistic manner, ie, the smaller the RMSE of an individual, the greater the probability of its selection.

### Mutation

The mutation operator of an evolutionary algorithm is supposed to ensure some amount of diversity within the population. In a classical implementation, it adds random noise to the chromosomes of selected individuals.

### Crossover

The crossover operator exchanges data between two selected parents and forms two new individuals and for that purpose, a standard 1-point crossover procedure can be used in which the cutting point is selected randomly.

### Offspring generation

At the end of each generation the merging elite, the mutated individuals and children created by the crossover operator creates offspring. The new population substitutes the previous one and the entire process is repeated until a satisfactory solution is found, or the maximum iteration reached.

## Experiment setup and results

To accomplish dimensionality reduction and identification of the corresponding regression model, the experiment was set up as follows: the dataset obtained for the PLGA dissolution profile had 300 features; therefore, the primary objective was to reduce the dimensions of the dataset. Hence, to accomplish this, the feature selection and feature extraction techniques discussed earlier were used. Subsequently, elementary prediction models were employed and their performances were assessed using ten-fold cross validation (10-CV) sets. Selection of the prediction model was based on the average of the RMSE computed over a set of ten results. In the final part of our experiment, we explored ensemble methods in order to exploit the elementary regression/prediction models.

## Feature selection method results

After cleaning and preprocessing the dataset, a feature selection treatment was used in which we used a backward feature elimination technique with the GPReg, LReg, SMOReg and REP prediction models. The parameter settings of the prediction models are provided in Table 2. The combination of attributes that offers the lowest RMSE was considered as the optimal feature set. For example, the optimal feature set obtained using the GPReg, LReg, MLP, SMOReg and REP regression models are 18, 32, 31, 30, and 31 with RMSE values of (resulting from a normalized dataset) 0.143, 0.156, 0.121, 0.153, and 0.126, respectively. The backward feature elimination results were convening in terms of RMSE. Therefore, for each of the predictors, we selected the feature sets with the smaller attributes, ie, set with ten, five, and one attribute.

We have stochastic feature selection techniques such as correlation-based, classifier-based, and wrapper-based methods. These feature selection methods were used to determine the merits (predictability) of different combinations of features. After assigning the merits of the several sets of features, the best first search (BFS) and the greedy search (greedy) methods were used to select the desired optimal feature set. Interestingly, in the present problem, when we used correlation-based feature selection, both the BFS and greedy searches produced identical feature sets with five attributes. The classifier-based feature selection was patched with GPReg, MLP, and LR eg, respectively, in order to evaluate the merits of the feature set. Subsequently, BFS and greedy searches were used to determine the optimal feature set. Therefore, we had class-GPReg-BFS, class-GPReg-greedy, class-MLP-BFS, class-MLP-greedy, class-LReg-BFS, and class-LReg-greedy feature selection methods, indicating a classifier-based method with GPReg as a feature set merit evaluator and BFS as the method to select the optimal feature set. Similarly, wrapper-GP-greedy, wrapper-MLP-greedy, and wrapper-LReg-greedy indicate a combination of wrapper-based feature selections, where GPReg, MLP, and LReg were used to evaluate the feature set. Interestingly, both BFS and greedy searches offered identical feature sets.

**Table 2** Parameters setting of the respective regression models used for the feature selection and feature extraction experiments

| Predictor | Parameters |
|---|---|
| GPReg | RBF kernel, gamma value = 1.0 |
| LReg | – |
| MLP | Three-layer MLP, hidden layer nodes - 50, learning rate - 0.3, momentum rate - 0.2 |
| SMOReg | Polynomial kernel, epsilon value - 0.001, tolerance level - 0.001 |
| REP tree | Max depth – no restriction |

**Abbreviations:** GPReg, Gaussian process regression; RBF, radial basis function; LReg, linear regression; MLP, multilayer perception; SMOReg, sequential minimal optimization regression; REP, reduced error pruning; –, no such parameter.

**Table 3** Experimental results for 10-CV datasets prepared with distinct random partitions of the complete dataset using feature selection technique (Identification of regression model)

| Selection method | Selected features | GPReg | LReg | MLP | REP | SMOReg |
|---|---|---|---|---|---|---|
| No selection | 300 | 16.81 | 17.07 | 18.57 | 13.05 | 17.95 |
| BFE | 1 | 27.47 | 26.61 | 28.33 | 24.37 | 26.97 |
| BFE | 5 | 17.11 | 23.45 | 23.11 | 14.23 | 23.38 |
| CFS | 5 | 20.80 | 25.08 | 22.41 | 18.31 | 25.42 |
| Class-MLP-greedy | 7 | 17.96 | 25.03 | 22.26 | 14.96 | 25.35 |
| BFE | 10 | 15.93 | 19.98 | 21.00 | 13.19 | 19.53 |
| Class-MLP-BFS | 15 | 15.88 | 22.90 | 16.83 | 13.91 | 24.23 |
| Wrapper-GPReg-greedy | 15 | 14.88 | 20.22 | 15.20 | 13.34 | 20.86 |
| Class-GPReg-BFS | 16 | 18.46 | 23.07 | 19.71 | 14.19 | 23.69 |
| Class-GPReg-greedy | 19 | 15.06 | 19.05 | 15.61 | 14.03 | 19.68 |
| Wrapper-MLP-greedy | 19 | 16.44 | 24.01 | 20.42 | 14.26 | 24.85 |
| Wrapper-LReg-greedy | 24 | 15.91 | 17.46 | 17.03 | 13.54 | 18.02 |
| BFE | Optimal* | 15.71 | 17.85 | 17.82 | 13.90 | 17.88 |
| Class-LReg-BFS | 31 | 15.95 | 16.92 | 15.63 | 14.00 | 17.58 |
| Class-LReg-greedy | 37 | 16.31 | 17.14 | 16.27 | 14.02 | 17.69 |

**Notes:** Values are the average of ten RMSE. *Optimal set of attributes for the GPReg, LReg, MLP, REP and SMOReg regression models are 18, 32, 31, 31, and 30, respectively.
**Abbreviations:** 10-CV, ten-fold cross-validation; GPReg, Gaussian process regression; LReg, linear regression; MLP, multilayer perception; REP, reduced error pruning; SMOReg, sequential minimal optimization; No, number; BFE, backward feature elimination; CFS, correlation-based feature selection; BFS, best fit search; wrapper, wrapper feature selection; greedy, greedy search; class, classifier-based feature selection.

A list of feature selection methods and the corresponding selected features are illustrated in Table 3.

## Results of the feature extraction technique

Unlike feature selection, feature extraction finds a new set of reduced features by computing linear or non-linear combinations of features from the available dataset. A comprehensive result is presented in Table 4, which illustrates the performance of feature extraction methods and regression models.

Dimensionality reduction tools offered by van der Maaten et al[20] were used for the feature extraction. PCA and FA linear dimensionality reduction methods, and non-linear dimensionality reduction methods such as kPCA and MDS were used to reduce the dimensions of the dataset from 300 to 50, 30, 20, 10, and 5. ICA was used to reduce the dimension of the dataset from 300 to 50. Results obtained using ICA are as follows. The mean RMSE and variance corresponding to GPReg, LReg, MLP, and SMOReg are 14.83, 17.23, 13.94, and 17.92 and 3.61, 2.34, 2.77, and 2.87, respectively. It may be observed from Table 4 that lower dimensions offer less significant improvement in terms of RMSE. However, if we compare the best results (the result of reducing the dimension to 50) of PCA (an RMSE of 13.59 corresponding to MLP) and ICA (an RMSE of 13.94 corresponding to MLP) with the result using all features (an RMSE of 16.812 corresponding to GPReg), it is evident that reducing the dimension significantly improves the performance of the prediction model. Examining Figure 2, an RMSE and variance comparison between chosen regression models applied on the dataset reduced it to a dimension of 50 using ICA, PCA, FA, kPCA, and MDS feature extraction techniques; we may conclude that the feature extraction using PCA performed best, both in terms of RMSE and variance, when the MLP regression model was used, whereas the feature extraction using ICA was second to PCA when MLP was used. When it came to GPReg, ICA had an edge over PCA.

## The regression model and ensemble results

In order to identify a suitable regression model, we chose several regression models. The parameter settings corresponding to the regression models are given in Table 3. A comprehensive feature selection result using 10-CV is presented in Table 3. Examining Table 3, we may therefore draw the following conclusions. First of all, in Table 3, we arranged the feature selection methods according to ascending order of the number of features selected by the feature selection methods; the first row of Table 3 that indicates no feature selection (ie, all 300 features were used), is exceptional. We compared the results of the prediction models arranged in the columns in Table 3. The feature selection process was able to find the most significant features that influenced the drug release rate. It may be observed that feature vectors from all the mentioned feature selection methods obtained a reduced

**Table 4** Experimental results for 10-CV datasets prepared with distinct random partitions of the complete dataset using feature extraction techniques

| Feature extraction method | | Regression model | Dimension reduction | | | | | | | | | |
|---|---|---|---|---|---|---|---|---|---|---|---|---|
| | | | 5 | | 10 | | 20 | | 30 | | 50 | |
| | | | Mean | VAR | Mean | VAR | Mean | VAR | Mean | VAR | Mean | VAR |
| Linear method | PCA | GPReg | 28.88 | 1.62 | 27.22 | 3.00 | 24.80 | 3.85 | 19.82 | 2.49 | 16.08 | 3.16 |
| | | LReg | 29.55 | 1.74 | 29.22 | 1.70 | 27.73 | 2.21 | 23.93 | 1.63 | 17.17 | 2.79 |
| | | MLP | 30.36 | 3.36 | 29.77 | 6.37 | 26.58 | 3.98 | 19.89 | 2.27 | 13.59 | 1.56 |
| | | SMOReg | 30.14 | 3.17 | 29.78 | 3.62 | 27.95 | 2.67 | 24.31 | 1.89 | 17.66 | 3.09 |
| | FA | GPReg | 29.23 | 1.77 | 28.56 | 2.67 | 28.31 | 3.34 | 28.30 | 3.42 | 28.26 | 3.31 |
| | | LReg | 29.97 | 1.77 | 29.97 | 1.77 | 29.97 | 1.77 | 29.97 | 1.77 | 29.98 | 1.82 |
| | | MLP | 30.64 | 2.02 | 30.50 | 1.91 | 31.01 | 1.83 | 30.93 | 2.30 | 30.91 | 0.77 |
| | | SMOReg | 30.28 | 3.45 | 30.28 | 3.45 | 30.26 | 3.37 | 30.29 | 3.44 | 30.28 | 3.46 |
| Non-linear method | Kernel PCA | GPReg | 28.60 | 1.68 | 27.08 | 2.12 | 24.96 | 1.96 | 24.32 | 2.17 | 22.81 | 4.43 |
| | | LReg | 29.31 | 1.52 | 28.05 | 1.78 | 25.35 | 2.05 | 25.17 | 2.23 | 22.98 | 4.27 |
| | | MLP | 29.81 | 3.57 | 29.65 | 7.94 | 27.07 | 4.09 | 25.97 | 5.52 | 25.27 | 8.49 |
| | | SMOReg | 29.43 | 1.41 | 28.68 | 1.65 | 25.90 | 1.70 | 25.79 | 2.00 | 23.24 | 4.76 |
| | MDS | GPReg | 28.91 | 2.17 | 28.73 | 2.47 | 28.41 | 3.16 | 28.24 | 3.17 | 28.16 | 3.27 |
| | | LReg | 29.56 | 1.86 | 29.21 | 2.08 | 29.19 | 2.08 | 29.11 | 1.92 | 29.14 | 2.04 |
| | | MLP | 30.42 | 3.71 | 29.38 | 4.11 | 29.93 | 3.10 | 30.01 | 4.53 | 29.98 | 4.42 |
| | | SMOReg | 29.98 | 2.62 | 29.64 | 2.55 | 29.64 | 2.76 | 29.66 | 2.85 | 29.65 | 2.89 |

**Note:** Mean and variance (VAR) is computed on ten RMSE obtained.
**Abbreviations:** 10-CV, ten-fold cross-validation; RMSE, root mean square error; PCA, principal component analysis; FA, factor analysis; MDS, multidimensional scaling; GPReg, Gaussian process regression; LReg, linear regression; MLP, multilayer perception; SMOReg, sequential minimal optimization regression.

set of the most influential features. Therefore, a general theory may be drawn about how and which features are the most dominant with regard to the PLGA drug release rate.

It is worth mentioning that the best result presented by Szlęk et al[21] is an RMSE of 15.4, considering eleven selected features using MLP and 17 features with an RMSE of 14.3 using MLP. From Table 3, it may be observed that when considering all 300 features, the best result we can achieve is by using REP, resulting in an RMSE of 13.05 (the average of the 10-CV result). Therefore, any regression model tested with a reduced feature set must compete with this result. In our study, the best result was obtained with the feature set using the wrapper-GPReg-greedy method with RSME of 14.88, 20.22, 15.20, 13.31, and 20.86 using the GPReg, LReg, MLP, REP and SMOReg elementary models, respectively. Therefore, we may consider the features "fused ring count", "heteroaromatic ring count", "largest ring system size", "chain atom count", "chain bond count", and "quaternary structure" from the protein descriptors group of features; "PVA concentration inner phase", "PVA concentration outer phase", "PVA molecular weight", and "PLGA to placticizer" from the formulation characteristics group of features; and "acetylsalicylic acid", "Szeged index", and "pH=12 logD" from the plasticizer group of features. From the emulsifier group, we have "a(yy)" and "time in days" as being the most influential feature sets obtained using wrapper-GPReg-greedy experiment. A complete list of the feature names can be found in Szlęk et al.[2]

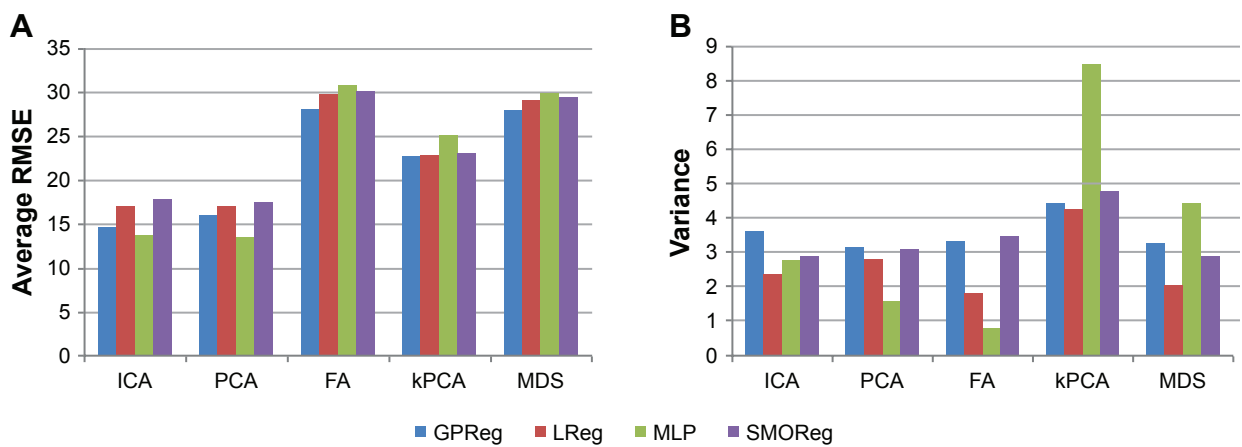

**Figure 2** Results of the feature extraction experiment for the reduced dimension set of 30 features: a comparison between the regression models. A comparison using average RMSE (**A**); a comparison using variances (**B**).
**Abbreviations:** RMSE, root mean square error; ICA, independent component analysis; PCA, principle component analysis; FA, factor analysis; kPCA, kernel PCA; MDS, multidimensional scaling; GPReg, Gaussian process regression; LReg, linear regression; MLP, multilayer perception; SMOReg, sequential minimal optimization regression.

After we obtained the best features, we resorted to using ensemble techniques. A comprehensive comparison of the results obtained using the ensemble methods and other elementary regression models is given in Table 5. From the results presented in Table 5, it is evident that some of the listed ensemble methods provides better results than that of the result produced by the best elementary predictor ie, reduced error pruning tree. The average RMSE obtained by ensemble such as RS using REP, RS using MLP, RS using GPReg, Bagging using REP, Bagging using MLP, mean output regression, quality weighted output regression, and EWE are 13.85, 18.20, 18.72, 11.49, 12.30, 10.43, 10.06, and 7.67, respectively.

## Discussion and analysis

In this article, experimental results obtained using both feature selection and feature extraction techniques are offered. The primary objective of the experiments was to find the lowest RMSE. In addition, we took advantage of the feature selection methods to obtain the best set of features. Our benchmark for the present experiment was the RMSE obtained using the complete set of features, ie, 300 features, and the results obtained by Szlęk et al.[21] The results obtained by the feature selection, feature extraction, and ensemble experiments are provided in Tables 3–5, respectively. The wrapper-based feature selection technique provided us the set of the most significant features. On the other hand, PCA offered a new set of features with solutions that were better than the solutions obtained with the complete dataset. The ensemble methods were only used for the feature selection methods. The ensemble methods enabled us to exploit all the evaluated regression models. Therefore, the best result (lowest RMSE) out of all the trained regressors was obtained using the EWE ensemble method. As mentioned above, predicting the PLGA dissolution rate is an important problem for the pharmaceutical industry. More significantly, identifying the influencing factors (features) is crucial for predicting the PLGA dissolution rate.

## Conclusion

Analyzing the effectiveness of the ensemble methods should be based on a comparison of the results obtained using the best elementary predictors. In our case, among the tested simple predictors, the lowest RMSE was reached with REP (13.34). The ensemble methods should improve regression accuracy over the best elementary predictor. The EWE ensemble method offered the lowest RMSE, which proves that in certain cases, combining the outputs of several predictors allows us to improve overall accuracy. It is essential to ensure diversity among the ensemble's constituents. Among the tested techniques, an ensemble of five heterogeneous regression algorithms provided the best results. Weighting their outputs was the most effective when weights were set

Table 5 A comprehensive conclusion of the results obtained from each regression model, including the ensemble techniques used

| Ensemble method | RMSE | Graphical representation |
|---|---|---|
| EWE | 7.67 | |
| QWOR | 10.06 | |
| MOR | 10.43 | |
| Bagging-REP | 11.49 | |
| Bagging-MLP | 12.30 | |
| RS-REP | 13.85 | |
| RS-MLP | 18.20 | |
| RS-GPReg | 18.72 | |
| SMOReg | 20.86 | |
| REP tree | 13.34 | |
| MLP | 15.20 | |
| LReg | 20.22 | |
| GPReg | 14.88 | |

**Note:** We have selected the feature set that was obtained using wrapper-GPReg-greedy search.
**Abbreviations:** RMSE, root mean square error; EWE, evolutionary weighted ensemble; QWOR, quality weighted output regression; MOR, mean output regression; bagging, ; MLP, multilayer perception; RS, random subspace; REP, reduced error pruning; SMOReg, sequential minimal optimization regression; LReg, linear regression; GPReg, Gaussian process regression.

using an evolutionary-based algorithm. Perhaps this is not the best method for creating a diversified ensemble of regression method in general, but it appeared to be the best one for the current problem we considered. We suggest that in all cases, a broad range of experiments with a variety of elementary regression algorithms and ensemble methods be used in order to find the best solution. Nonetheless, the obtained results prove that the proposed EWE method is an effective option for finding a solution to the present problem.

## Acknowledgments

This work was supported by the IPROCOM Marie Curie Initial Training Network, funded through the People Programme (Marie Curie Actions) of the European Union's Seventh Framework Programme FP7/2007–2013/, under REA grant agreement number 316555. This work was also supported by the Polish National Science Center under grant number DEC-2013/09/B/ST6/02264.

## Disclosure

The authors report no conflicts of interest in this work.

Dovepress

Function approximation of PLGA micro and nanoparticle dissolution rate34. Hall MA, Smith LA. Practical feature subset selection for machine learning. In: McDonald C, editor. *Proceedings of the 21st Australasian Computer Science Conference ACSC '98, Perth, WA, Australia, 4–6 February, 1998.* Berlin: Springer, 1998:181–191.
35. Kohavi R, John GH. Wrappers for feature subset selection. *Artif Intell.* 1997;97(1):273–324.
36. Pearson K. LIII. On lines and planes of closest fit to systems of points in space. *Lond Edinburgh Dublin Philosoph Mag J Sci.* 1901;2(11):559–572.
37. Abdi H, Williams LJ. Principal component analysis. *Wiley Interdisciplinary Reviews: Computational Statistics.* 2010;2(4):433–459.
38. Harman HH. *Modern Factor Analysis.* Chicago: University of Chicago Press; 1960.
39. Hyvärinen A, Oja E. Independent component analysis: algorithms and applications. *Neural Netw.* 2000;13(4–5):411–430.
40. Hyvärinen A. Fast and robust fixed-point algorithms for independent component analysis. *IEEE Trans Neural Netw.* 1999;10(3):626–634.
41. Hyvärinen A. Independent component analysis and blind source separation. Technical report. Helsinki: Helsinki University of Technology; 2003. Available from: http://www.cs.helsinki.fi/u/ahyvarin/presentations/Berlin05.pdf. Accessed December 22, 2014.
42. Schölkopf B, Smola A, Müller KR. Kernel principal component analysis. In: Gersner W, Germond A, Hasler M, Nicoud JD, editors. *Artificial Neural Networks — ICANN '97.* Berlin: Springer; 1997:583–588.
43. Kruskal JB. Multidimensional scaling by optimizing goodness of fit to a nonmetric hypothesis. *Psychometrika.* 1964;29(1):1–27.
44. Neter J, editor. *Applied linear statistical models.* 4th ed. Chicago: Irwin; 1996.
45. Rasmussen CE, Williams CK. *Gaussian Processes for Machine Learning (Adaptive Computation and Machine Learning).* Cambridge, MA: MIT Press; 2005.
46. Rasmussen CE, Nickisch H. Gaussian processes for machine learning (GPML) toolbox. *JMLR.* 2010;11:3011–3015.
47. Haykin S. *Neural Networks: A Comprehensive Foundation.* 1st ed. Upper Saddle River, NJ: Prentice Hall PRT; 1994.
48. Werbos PJ. *Beyond Regression: New Tools for Prediction and Analysis in the Behavioral Sciences.* Princeton, NJ: Harvard University Press; 1975.
49. Rumelhart DE, McClelland JL, editors. *Parallel Distributed Processing: Explorations in the Microstructure of Cognition, Vol 1: Foundations.* Cambridge, MA: MIT Press; 1986.
50. Riedmiller M, Braun H. A direct adaptive method for faster back-propagation learning: the RPROP algorithm. In: *IEEE International Conference on Neural Networks.* IEEE, 1993:586–591.
51. Quinlan JR. Simplifying decision trees. *Int J Man-Mach Stud.* 1987;27(3):221–234.
52. Mohamed WN, Salleh MN, Omar AH. A comparative study of reduced error pruning method in decision tree algorithms. In: *IEEE International Conference on Control System, Computing and Engineering (ICCSCE)*; November 23-25, 2012; Penang, Malaysia. November 2012. 2012;392–397.
53. Smola AJ, Schölkopf B. *Learning with Kernels: Support Vector Machines, Regularization, Optimization, and Beyond.* Cambridge, MA: MIT Press; 1998.
54. Smola AJ, Schölkopf B. A tutorial on support vector regression. *Statistics Computing.* 2004;14(3):199–222.
55. Schölkopf B, Burges CJC, Smola AJ. *Advances in Kernel Methods: Support Vector Learning.* Cambridge, MA: MIT Press; 1999.
56. Platt J. Probabilistic outputs for support vector machines and comparisons to regularized likelihood methods. *Advances in Large Margin Classifiers.* 1999;10(3):61–74.
57. Brown G, Wyatt J, Harris R, Yao X. Diversity creation methods: A survey and categorisation. *Journal of Information Fusion.* 2005;6(1):5–20.
58. Ho TK. The random subspace method for constructing decision forests. *IEEE Transac Pattern Anal Mach Intell.* 1998;20(8):832–844.
59. Breiman L. Bagging predictors. *Mach Learn.* 1996;24(2):123–140.
60. Saeys Y, Inza I, Larrañaga P. A review of feature selection techniques in bioinformatics. *Bioinformatics.* 2007;23(19):2507–2517.
61. Bauer E, Kohavi R. An empirical comparison of voting classification algorithms: Bagging, boosting, and variants. *Mach Learn.* 1999;36(1–2):105–139.
62. Goldberg DE. *Genetic Algorithms in Search, Optimization, and Machine Learning.* Boston, MA: Addison-Wesley; 1989.
**International Journal of Nanomedicine**

**Publish your work in this journal**

The International Journal of Nanomedicine is an international, peer-reviewed journal focusing on the application of nanotechnology in diagnostics, therapeutics, and drug delivery systems throughout the biomedical field. This journal is indexed on PubMed Central, MedLine, CAS, SciSearch®, Current Contents®/Clinical Medicine, Journal Citation Reports/Science Edition, EMBase, Scopus and the Elsevier Bibliographic databases. The manuscript management system is completely online and includes a very quick and fair peer-review system, which is all easy to use. Visit http://www.dovepress.com/testimonials.php to read real quotes from published authors.

**Submit your manuscript here:** http://www.dovepress.com/international-journal-of-nanomedicine-journal

Dovepress
International Journal of Nanomedicine 2015:10

submit your manuscript | www.dovepress.com

**1129**

Dovepress